\documentclass[aps, prb, twocolumn,superscriptaddress]{revtex4-1}

\usepackage{graphicx}
\usepackage{color}

\newcommand{\cel}[1]{\ensuremath{#1\,^\circ\textrm{C}}}

\begin{document}
\title{Ferromagnetism or slow paramagnetic relaxation in Fe-doped Li$_3$N?}

\author{M. Fix}
 \affiliation{EP VI, Center for Electronic Correlations and Magnetism, Institute of Physics, University of Augsburg, D-86159 Augsburg, Germany}
\author{A. Jesche}
 \email[]{anton.jesche@physik.uni-augsburg.de}
 \affiliation{EP VI, Center for Electronic Correlations and Magnetism, Institute of Physics, University of Augsburg, D-86159 Augsburg, Germany}
 \affiliation{The Ames Laboratory, Iowa State University, Ames, Iowa 50011, USA}
\author{S. G. Jantz}
 \affiliation{Chair of Solid State Chemistry, Institute of Physics, University of Augsburg, D-86159 Augsburg, Germany}
\author{S. A. Br\"auninger}
 \affiliation{Institute for Solid State and Materials Physics, TU Dresden, D-01069 Dresden, Germany}
\author{H.-H. Klauss}
 \affiliation{Institute for Solid State and Materials Physics, TU Dresden, D-01069 Dresden, Germany}
\author{R. S. Manna}
 \affiliation{EP VI, Center for Electronic Correlations and Magnetism, Institute of Physics, University of Augsburg, D-86159 Augsburg, Germany}
 \affiliation{Department of Physics, IIT Tirupati, Tirupati 517506, India}
\author{I. M. Pietsch}
 \affiliation{EP VI, Center for Electronic Correlations and Magnetism, Institute of Physics, University of Augsburg, D-86159 Augsburg, Germany}
\author{H. A. H\"oppe}
 \affiliation{Chair of Solid State Chemistry, Institute of Physics, University of Augsburg, D-86159 Augsburg, Germany}
\author{P. C. Canfield}
 \affiliation{The Ames Laboratory, Iowa State University, Ames, Iowa 50011, USA}
 \affiliation{Department of Physics and Astronomy, Iowa State University, Ames, Iowa 50011, USA}

\begin{abstract}
We report on isothermal magnetization, M\"ossbauer spectroscopy, and magnetostriction as well as temperature-dependent alternating-current (ac) susceptibility, specific heat, and thermal expansion of single crystalline and polycrstalline Li$_2$(Li$_{1-x}$Fe$_x$)N with $x = 0$ and $x \approx 0.30$. 
Magnetic hysteresis emerges at temperatures below $T \approx 50$\,K with coercivity fields of up to $\mu_0H = 11.6$\,T at $T = 2$\,K and magnetic anisotropy energies of 310\,K (27\,meV).
The ac susceptibility is strongly frequency dependent ($f$\,=\,10--10,000\,Hz) and reveals an effective energy barrier for spin reversal of $\Delta E \approx 1100$\,K. 
The relaxation times follow Arrhenius behavior for $T > 25$\,K.
For $T < 10$\,K, however, the relaxation times of $\tau \approx 10^{10}$\,s are only weakly temperature-dependent indicating the relevance of a quantum tunneling process instead of thermal excitations. 
The magnetic entropy amounts to more than 25\,J mol$^{-1}_{\rm Fe}$\,K$^{-1}$ which significantly exceeds $R$ln2, the value expected for the entropy of a ground state doublet.
Thermal expansion and magnetostriction indicate a weak magneto-elastic coupling in accordance with slow relaxation of the magnetization.
The classification of Li$_2$(Li$_{1-x}$Fe$_x$)N as ferromagnet is stressed and contrasted with highly anisotropic and slowly relaxing paramagnetic behavior. 
\end{abstract}

\maketitle

\section{Introduction}
The stability of magnetic states is of vital importance on various length scales ranging from nanometer size bits in magnetic recording media\,\cite{Plumer2011} to bulk hard magnets in electro motors and wind turbines\,\cite{Gutfleisch2011}. 
Two basic ingredients for stable, remnant magnetization are strong interaction and sufficient magnetic anisotropy: ferromagnetic interactions cause a parallel orientation of spins in competition with thermal disorder, magnetic anisotropy leads to a preferred alignment along a certain direction. 
Without anisotropy, the total magnetization of a ferromagnet drops from full saturation to zero as soon as an external field is removed (soft ferromagnet). 
But is it possible to remain a spontaneous magnetization purely based on anisotropy without invoking ferromagnetic interactions?
An arbitrary large anisotropy seems to prohibit at least the combination of reaching saturation (in limited fields) and maintaining a remnant magnetization: thermal excitations prevent saturation at higher temperatures. 
At low temperatures, on the other hand, it is impossible to overcome the energy barrier associated with the large anisotropy energy by applying a finite field. 

Li$_2$(Li$_{1-x}$Fe$_x$)N\,\cite{Klatyk1999} with $x \approx 0.30$ shows among the largest magnetic anisotropy and coercivtiy known\,\cite{Jesche2014b, Ke2015}. 
The anisotropy energy was estimated to 27\,meV\,\cite{Jesche2015, Xu2017}, the anisotropy field extrapolates to 220\,T. 
The magnetic easy axis is oriented along the crystallographic c-axis and saturation at $T = 10$\,K is obtained in an applied field of a few Tesla (see below). 
Diluted samples with $x < 1$\,\% contain predominantly isolated, non-interacting Fe moments and show a significant time-dependence of the magnetization\,\cite{Jesche2014b} that is driven by quantum tunneling of the magnetization\,\cite{Sessoli2003}.
For larger Fe-concentrations with $x \approx 0.30$, however, those time-dependencies seemed much less pronounced and no steps in the isothermal magnetization loops that would indicate quantum tunneling were found\,\cite{Klatyk2002, Jesche2014b}. 
Based on temperature-dependent and isothermal magnetization measurements\,\cite{Klatyk2002, Jesche2014b} as well as M\"ossbauer spectroscopy\,\cite{Klatyk2002, Ksenofontov2003} those samples could be considered 'classical' ferromagnetic materials with a Curie temperature of $T_{\rm C} \sim 65$\,K. 

Here we show that pronounced time-dependence of the magnetization of Li$_2$(Li$_{1-x}$Fe$_x$)N is observed for all iron concentrations $x$ with slower relaxation for larger $x$. 
The outline of the paper is as follows: in Sec.\,\ref{2exp} and \,\ref{3growth} we describe experimental details, crystal growth process and structural characterization. 
Isothermal magnetization, time-dependent magnetization and alternating current (ac) magnetic susceptibility are presented in Sec.\,\ref{ch:magnetic_properties}, followed by M\"ossbauer spectroscopy in Sec.\,\ref{moess}.
Specific heat is shown in Sec.\,\ref{5heat}, thermal expansion and magnetostriction in Sec.\,\ref{6dila}.
In Sec.\,\ref{7disc} the results are discussed and contrasted with the behavior observed in spin-glasses, superparamagnets and molecular magnets. 

\section{Experimental}\label{2exp}

X-Ray powder diffraction (XRPD) was performed at room temperature on ground single crystals using a Bruker D8 Advance powder diffractometer (Cu K$\alpha_1$ radiation). To prevent sample degradation the glass capillaries used for XRPD were prepared under argon atmosphere. Lattice parameters were obtained from Rietveld refinements using GSAS\,\cite{Larson2000} and EXPGUI\,\cite{Toby2001}:
Instrument parameters for profile function 2 were determined prior to
the measurement using a Si-standard. Only lattice parameters, sample
displacement, transparency, Lorentzian coefficients and isotropic displacement parameters were released - all other parameters (except background and scaling)
were kept constant during the refinement. In this way, weighted
profile R-factors of $R_\mathrm{wp} = 1.5, 3.8~{\rm  and}~ 4.4$ were achieved for Fe-doped, as-grown and annealed Li$_3$N, respectively (restricted to the Bragg contribution to the diffraction pattern).
Laue back reflection patterns were taken with a digital Dual FDI NTX camera manufactured by Photonic Science (tungsten anode, $U = 15$\,kV).

Chemical analysis was performed using inductively coupled plasma mass spectrometry (ICP-MS, Thermo Scientific, samples used for magnetization and specific heat measurements) and inductively coupled plasma optical emission spectroscopy (ICP-OES, Vista-MPX, samples used for XRPD and thermal expansion measurements). To this end the samples were dissolved in a mixture of hydrochloric acid and distilled water. 

Isothermal magnetization and ac susceptibility were measured using a Quantum Design Physical Property Measurement System (PPMS), equipped with a 14\,T magnet.
Time dependent magnetization data were obtained using a 7\,T Magnetic Property Measurement System (MPMS), manufactured by Quantum Design.

The M\"ossbauer measurements were performed using a standard WissEL spectrometer in transmission geometry employing a $^{57}$Co source with an initial activity of 1.4\,GBq.
The drive was run in sinusoidal mode minimizing the velocity error.
The measurements were carried out in a Cryovac helium flow cryostat.
The M\"ossbauer spectra were analyzed with the {\it Moessfit} software package\,\cite{Kamusella2016} using transmission integral simulation.
Isomer shifts are given relatively to room temperature $\alpha$-Fe.

Specific heat measurements were carried out with a heat-pulse relaxation method using a Quantum Design PPMS.
Thermal expansion and magnetostriction measurements were performed using a high-resolution capacitive dilatometer placed inside a Quantum Design 14\,T PPMS\,\cite{Pott1983, Kuchler2012}.

\section{Crystal growth and structural characterization}\label{3growth}

\begin{figure}
\centering
\includegraphics[width=.47\textwidth]{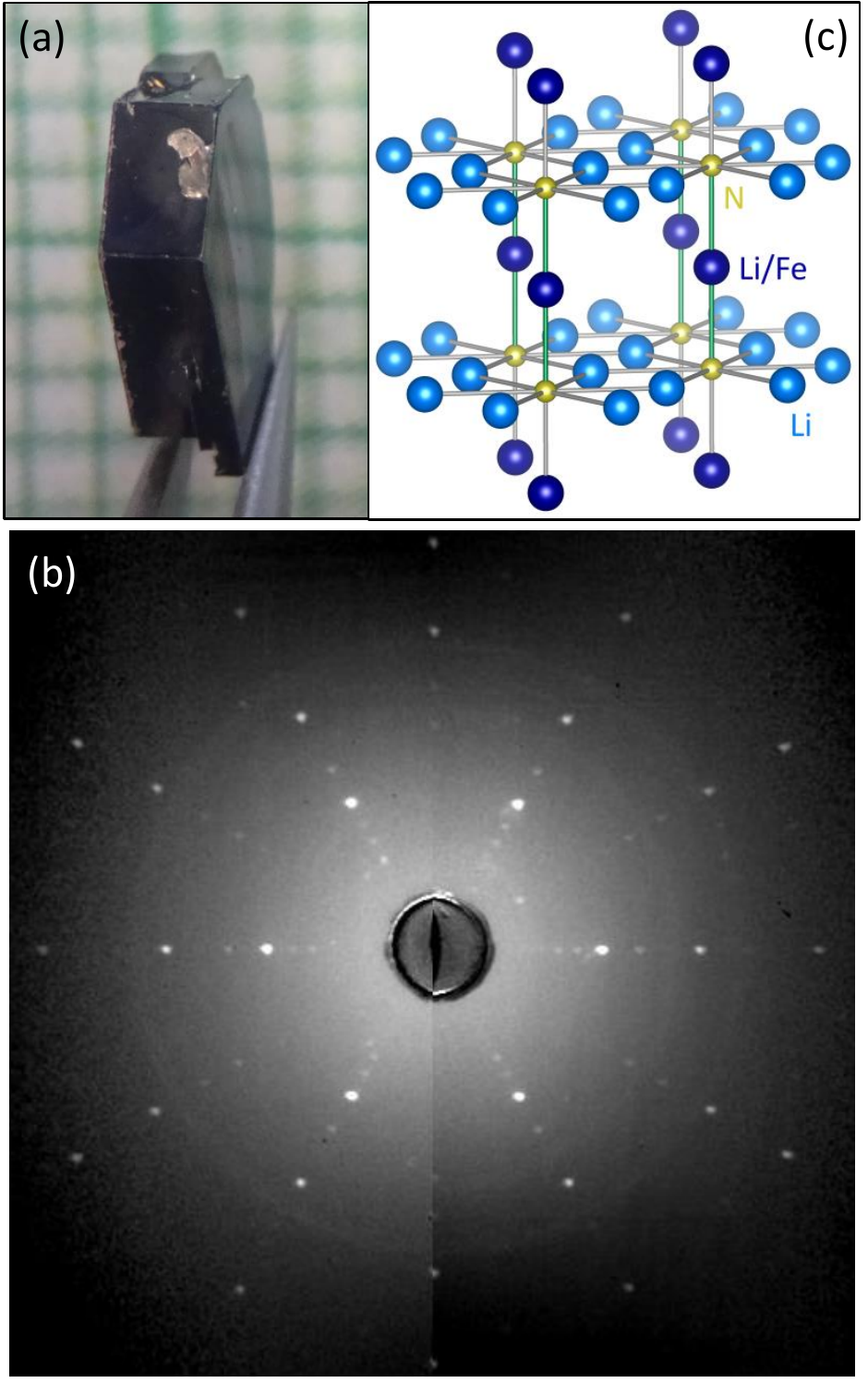}
\caption{(a) Single crystal of Li$_3$N on a millimeter grid and (b) corresponding Laue-back-reflection pattern. (c) Crystal structure of Li$_2$(Li$_{1-x}$Fe$_x$)N comprising two different Li sites, one of which is partially occupied by Fe.}
\label{growth}
\end{figure}

The crystals were grown out of a lithium rich solution. The doped and undoped samples were grown under similar conditions: The starting materials Li$_3$N powder (Alfa Aesar, 99.4\,\%), Li granules (Alfa Aesar, 99\,\%) and Fe granules (Alfa Aesar 99.98\,\%) were mixed in a molar ratio of Li:Fe:Li$_3$N = 9:0:1 and 8.7:0.3:1 for the undoped and the doped samples, respectively. 
The mixtures with a total mass of roughly 1.5\,g were packed into a three-cap Ta crucible\, \cite{Canfield2001, Jesche2014c} inside an Argon-filled glovebox. The crucibles were sealed in $\sim 0.6$\,bar Ar via arc welding and finally sealed in a silica ampule in $\sim 0.1$\,bar Ar. 
The mixtures were heated to $T$ = \cel{900} within 5\,h, cooled to $T$ = \cel{750} over 1.5\,h, slowly cooled to $T$ = \cel{500} over 60\,h and finally decanted to separate the crystals from the excess flux. 
Figure\,\ref{growth}a shows a representative picture of the obtained single crystals.
A Laue-back-reflection pattern of a Li$_3$N single crystal, confirming the sixfold crystallographic c-axis to be perpendicular to the larger surface of the obtained plate-like crystals, is shown in Fig.\,\ref{growth}b.

\begin{figure}
 \includegraphics[width=0.48\textwidth]{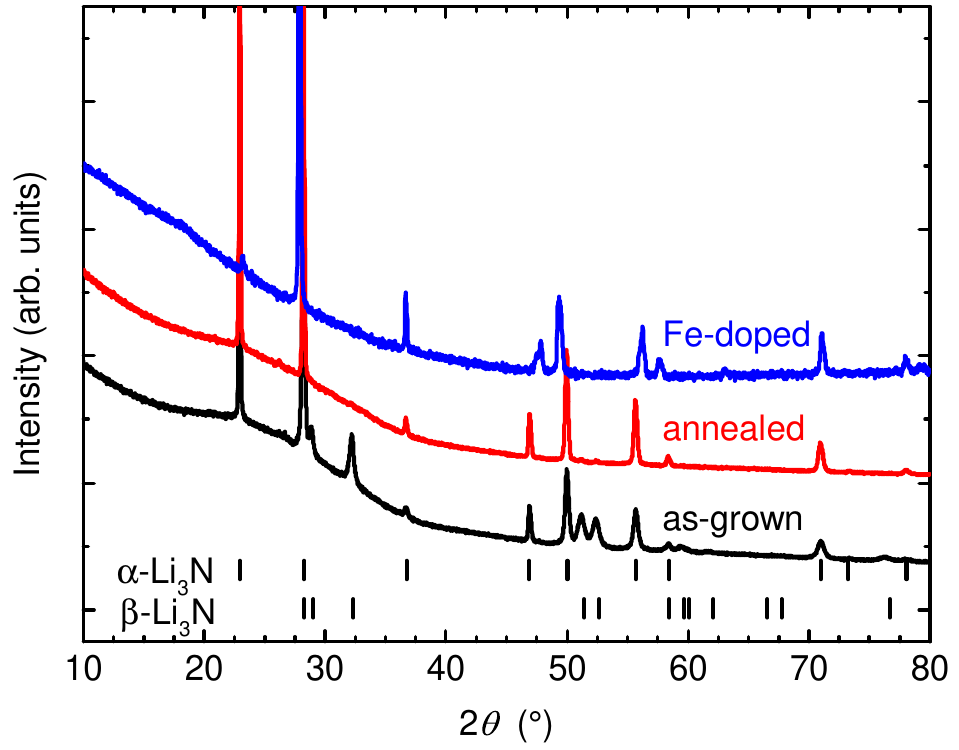}
 \caption{XRPD pattern of Li$_2$(Li$_{1-x}$Fe$_x$)N with $x = 0$ and $x = 0.25$. 
Li$_3$N after grinding (bottom, black curve) and subsequent annealing (middle, red curve). 
Theoretical peak positions of $\alpha$ and $\beta$ phase are indicated by tick marks\,\cite{David2007}. The pattern obtained for $x = 0.25$ is shown by the upper blue curve. 
Whereas the pressure applied upon grinding partially induces a phase transition from the $\alpha$ to the $\beta$ phase in undoped Li$_3$N, no such transformation is observed for $x = 0.25$.}
 \label{xrpd}
\end{figure}

Li$_2$(Li$_{1-x}$Fe$_x$)N crystallizes in a hexagonal lattice [space group $P\,6/m\,m\,m$ (191)]. The substituted Fe atoms occupy only the Li-1$b$
Wyckoff position\,\cite{Klatyk1999}, leading to a linear, two-fold coordination of the transition metal between two nitrogen ions (Fig.\,\ref{growth}c). 
The XRPD data measured on ground single crystals with $x = 0$ and $x = 0.25$ are plotted in Fig.\,\ref{xrpd}.
For the as-grown (ground) sample with $x = 0$ additional Bragg peaks are observed that can be indexed based on the known high-pressure phase $\beta$-Li$_3$N [space group P6$_3$/\textit{mmc} (194)]\,\cite{Beister1988}. 
Since the critical pressure of $\sim0.6$\,GPa is rather low, the partial transformation of the $\alpha$ phase of Li$_3$N to the $\beta$ phase takes place upon grinding the single crystals for XRPD\,\cite{Beister1988}. 
The as-grown single crystals, on the other hand, consist of only $\alpha$-Li$_3$N, as confirmed by the Laue-back-reflection pattern (Fig.\ref{growth}b).

Annealing of the powder at ambient pressure at temperatures above \cel{200} leads to a re-transformation of the $\beta$ to the $\alpha$ phase\,\cite{Beister1988,Huq2007}. 
In accordance, the peaks associated with $\beta$-Li$_3$N disappeared after annealing the powder for 12\,h at \cel{267} in the sealed glass capillary (red pattern plotted the middle of Fig.\,\ref{xrpd}).
For the Fe-doped samples no transformation to the $\beta$ phase was observed (blue pattern plotted at the top of Fig.\,\ref{xrpd}). This is probably due to the lower pressure required for grinding these (more brittle) samples.
Accordingly, all physical properties presented in this publication refer to the $\alpha$ phase.

The obtained lattice parameters are summarized in Tab.\,\ref{tab:latticeparameters}. 
The values of $a = 365.1$\,pm and $c = 387.2$\,pm obtained for undoped $\alpha$-Li$_3$N are in good agreement with previous results (e.g.: $a=364.8$\,pm and $c=387.5$\,pm\,\cite{Beister1988} or $a = 365.3$\,pm and $c = 387.4$\,pm\,\cite{David2007}. 
For Li$_3$N with $25$\,\% Fe substitution, as determined by ICP-OES, an increase of $a$ by 1.1\,\% ($a = 369.0$\,pm) and a decrease of $c$ by 1.7\,\% ($c = 381.2$\,pm) are found in reasonable agreement with Refs.\,\onlinecite{Yamada2011, Jesche2014b}.

\begin{table}
	\begin{tabular}{l c c}
		\hline
		\hline		
		~ & $a$\,(pm) & $c$\,(pm)\\
		\cline{2 -3}
		as-grown $\alpha$-Li$_3$N & ~365.2~ & ~387.9~\\
		annealed $\alpha$-Li$_3$N & 365.1 & 387.2\\
		Fe-doped $\alpha$-Li$_3$N & 369.0 & 380.9\\
		\hline
		as-grown $\beta$-Li$_3$N \phantom{---}& 357.3 & 635.2\\
		\hline
		\hline
	\end{tabular}
	\caption{Lattice parameters $a$ and $c$ of as-grown and annealed Li$_3$N in comparison with the Fe-doped sample ($x = 0.25$). The values were determined by Rietveld refinement of the data shown in Fig \ref{xrpd}.}
	\label{tab:latticeparameters}
\end{table}

\section{Magnetic properties}
\label{ch:magnetic_properties}
Effective magnetic moments of $\mu_{\rm eff}^{\perp c} = 4.6\,\mu_{\rm B}$ and $\mu_{\rm eff}^{\parallel c} = 6.3(4)\,\mu_{\rm B}$ for $x = 0.28$ were reported in a previous publication\,\cite{Jesche2014b}. 
For a polycrystalline, cold-pressed pellet with $x = 0.29$ we do also observe Curie-Weiss behavior with an effective moment of 4.9\,$\mu_{\rm B}$ per iron (not shown).
In the following we report on detailed measurements of low-temperature isothermal magnetization, ac-susceptibility and direct time-dependence of the magnetization measured on polycrystalline and single crystalline samples.
All measurements shown in this section were performed on the same samples: a single crystal with $x = 0.28$ ($m = 8.9$\,mg) and a cold-pressed, polycrystalline pellet with $x = 0.29$ ($m = 17.1$\,mg).

\subsection{Isothermal magnetization}
\begin{figure*}
 \includegraphics[width=0.9\textwidth]{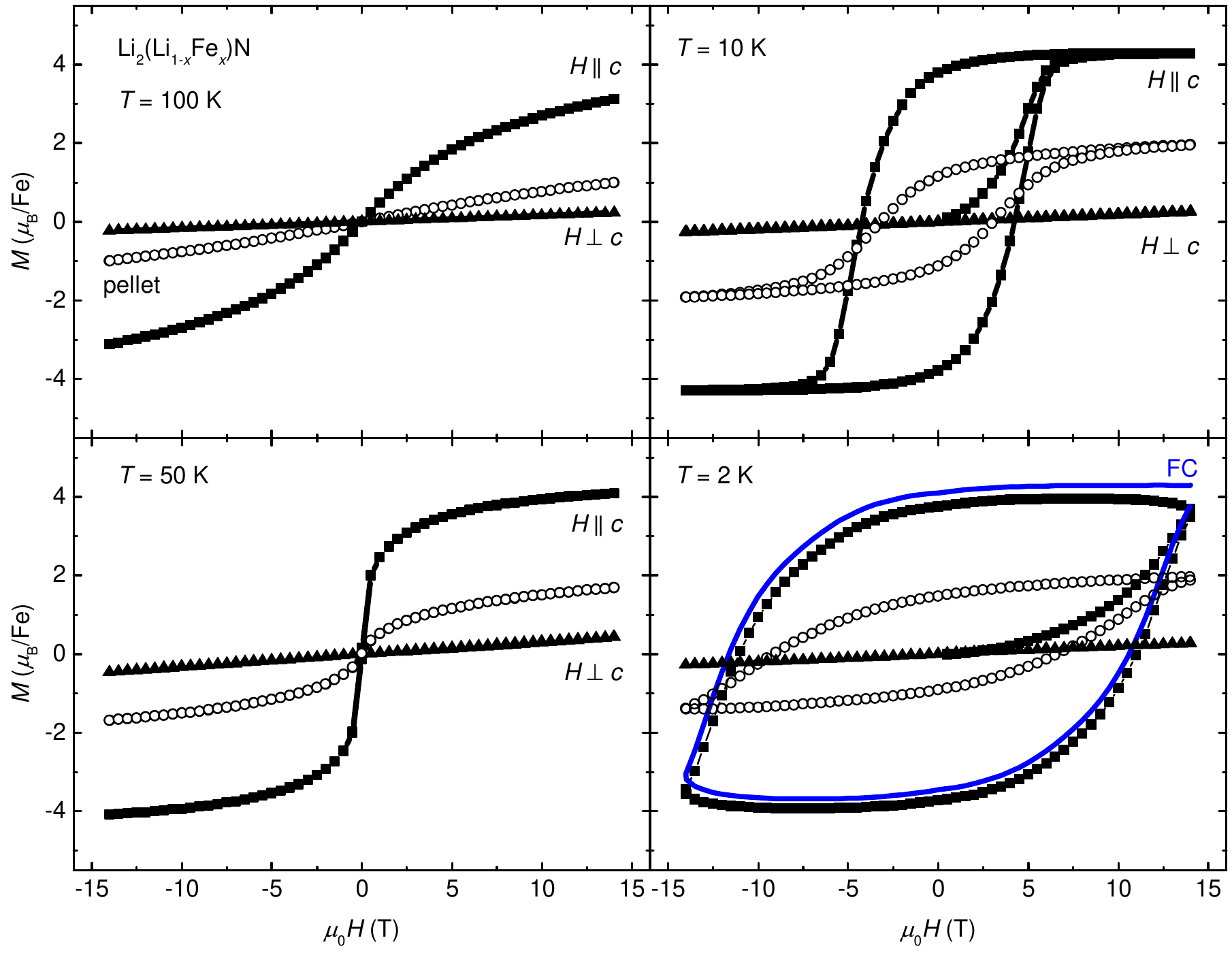}
 \caption{Isothermal magnetization of Li$_2$(Li$_{1-x}$Fe$_x$)N measured on a polycrystalline pellet ($x = 0.29$, open circles) and on a single crystal ($x = 0.28$, closed symbols) at the given temperatures. Hysteresis emerges for $T < 50$\,K with coercivity fields of up to $\mu_0 H \sim 11.6$\,T. The blue, solid line in the bottom panel shows a field-cooled (FC) measurement of the single crystal ($H \parallel c$).}
 \label{m-h}
\end{figure*}

Isothermal magnetization, $M$-$H$, of Li$_2$(Li$_{1-x}$Fe$_x$)N measured at different temperatures are plotted in Fig.\,\ref{m-h}.
The effective sweep rate for the full loops was 3.7\,mT/s (10\,mT/s between the measurements). 
The magnetization shows a pronounced anisotropy over the whole temperature range investigated ($T = 2$--100\,K). 
Whereas saturation with a large moment of $\mu_\mathrm{sat}^{\parallel c} \approx 4.3\,\mu_\mathrm{B}$ is approached for magnetic fields applied parallel to the crystallographic c-axis, $H \parallel c$, the magnetization is much smaller for $H \perp c$ and increases only slowly with field.
Even the largest available field of $\mu_0H = 14$\,T is not sufficient to reach saturation at $T = 2$\,K from a ZFC state ($H \parallel c$; saturation is observed in FC measurements, see solid blue line in the bottom panel of Fig.\,\ref{m-h}).
The anisotropy energy estimated from the extrapolated anisotropy field of 220\,T amounts to 27\,meV (310\,K). 

For temperatures $T < 50$\,K magnetic hysteresis becomes apparent for $H \parallel c$ with a large coercivity field of $\mu_0 H_\mathrm{c} = 11.6\,\mathrm{T}$ at $T = 2\,\mathrm{K}$.
In the polycrystalline pellet the hysteresis is somewhat smaller with a coercivity field of $\mu_0 H_\mathrm{c} = 9.2\,\mathrm{T}$ at $T = 2$\,K.

\subsection{Time-dependent magnetization}
\begin{figure}
 \includegraphics[width=0.47\textwidth]{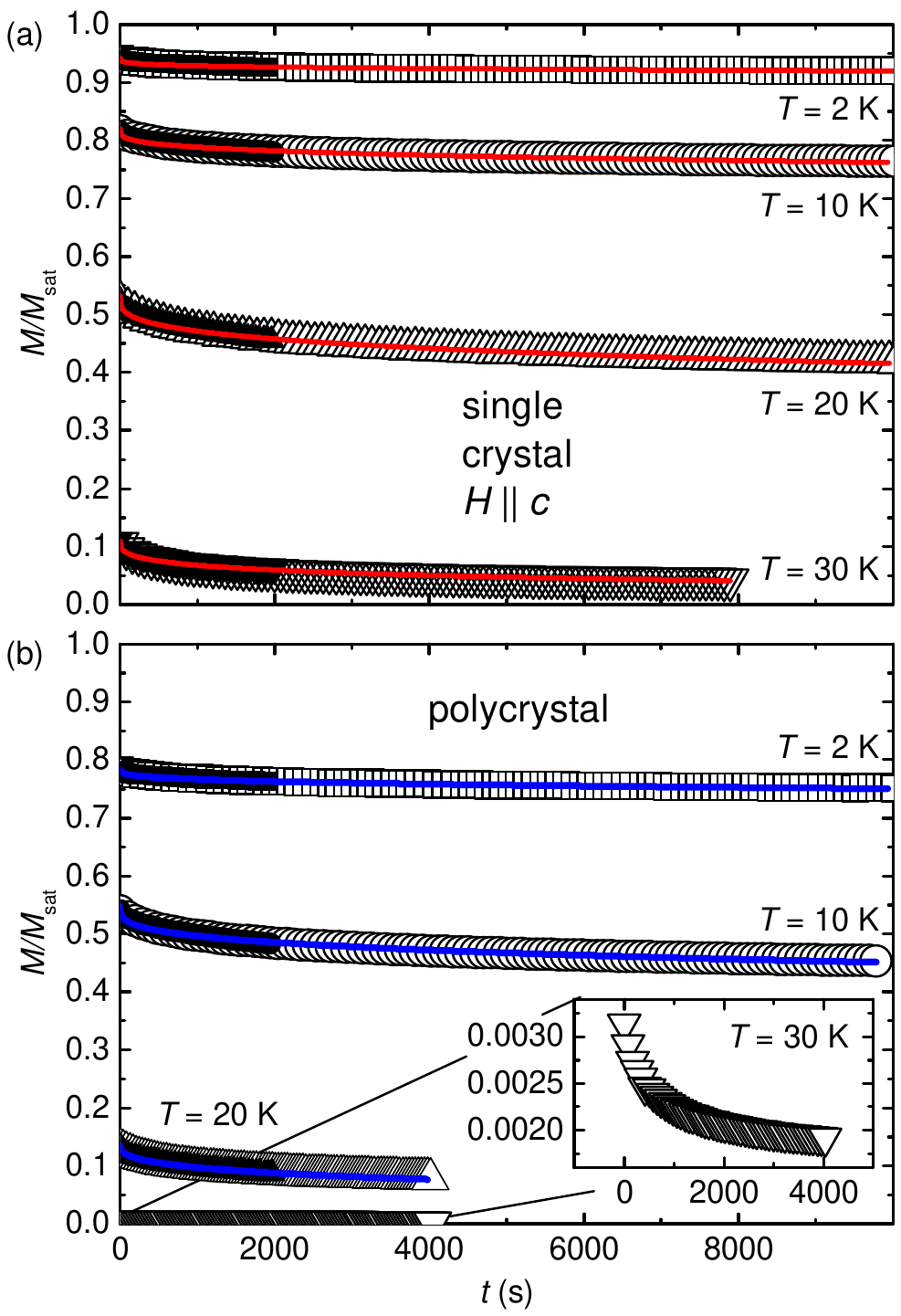}
 \caption{Time-dependence of the magnetization of Li$_2$(Li$_{1-x}$Fe$_x$)N in zero-field at constant temperature. After cooling in $\mu_0H = 7$\,T to the temperatures given, the applied field was ramped to $H = 0$. Experimental data are given by the larger, open symbols. 
(a) single crystal with $x = 0.28$ and $H \parallel c$. 
(b) polycrystalline, cold-pressed pellet with $x = 0.29$. 
Fits to a stretched exponential function fail to describe the time-dependence of the full range with good accuracy (solid red/blue lines). 
A critical estimate of the error of the obtained relaxation time, $\Delta \tau$, was performed by restricting the fits to the beginning (smaller, filled, black symbols) and the latter part of the decay.}
 \label{time}
\end{figure}

The time-dependence of the magnetization $M(t)$ of Li$_2$(Li$_{1-x}$Fe$_x$)N at constant temperature and zero-field is shown in Fig.\,\ref{time}.
$M(t)$ curves were recorded at temperatures $T$\,=\,2--30\,K. 
Prior to the measurement the samples were cooled in an applied field $\mu_0 H = 7$\,T (FC) to the temperature given in the plot ($H \parallel c$ for the single crystal). 
At that point, the magnetization was saturated (or close to saturation: for the highest temperature of $T = 30$\,K, 98\,\%  and 89\,\% of $M_{\rm sat}$ were observed for single crystal and pellet, respectively).
Afterwards, the field was ramped to $H = 0$ at the maximum rate (that is within $470\,{\rm s}\pm10$\,s).

In order to quantify the time-dependence, a stretched exponential function 
\begin{equation}
M(t) = M_0 \exp\left[ - \left( t/\tau \right) ^\beta \right],
\label{eq:timedependence}
\end{equation}
	
with the relaxation time $\tau$ and  $M_0 = M(t=0)$ was fit to the data ($t=0$ was determined as the moment when the magnetic field was stabilized).
The exponent $\beta$ accounts for a deviation from exponential decay, for example due to a change of the internal field during the relaxation process (such fits are frequently used to describe $M(t)$ in molecular magnets, see e.g. Refs.\,\onlinecite{Sangregorio1997,Sessoli2003}).
However, the fits fail to describe the full range of the measured $M(t)$ with good accuracy (red and blue solid lines in Fig.\,\ref{time}). 
The deviation of $M(t)$ from a simple exponential decay is too strong to be described by the exponent $\beta$.
Nevertheless, as we will see below, the observed $\tau$ shows a clear dependence on temperature in good agreement with the values obtained from ac-susceptibility. 

In order to estimate the error of the relaxation time, $\Delta \tau$, additional fits were performed that were restricted to the beginning ($t = 0$--2000\,s, see filled, smaller symbols in Fig.\ref{time}) and the latter part of the decay ($t = 5000$--10000\,s). 
The former yields the lower limit of $\tau$ (faster relaxation at short times), the latter the upper limit, giving rise to asymmetric error bars. 
Exponents of $\beta_{\rm sc} = 0.27$--0.34 and $\beta_{\rm pellet} = 0.31$--0.40 (with a clear tendency to increase with temperature) were found for fitting the full range of $M(t)$ for single crystal and pellet, respectively. 
For the fits restricted to short times, a slightly larger and temperature independent value of $\beta \approx 0.5$ was found for both samples. 
In order to ensure convergence for the fit of the latter part of the decay ($t = 5000$--10000\,s), the exponent had to be fixed (to the value obtained for fitting the full range). 
 
\begin{figure}
 \includegraphics[width=0.47\textwidth]{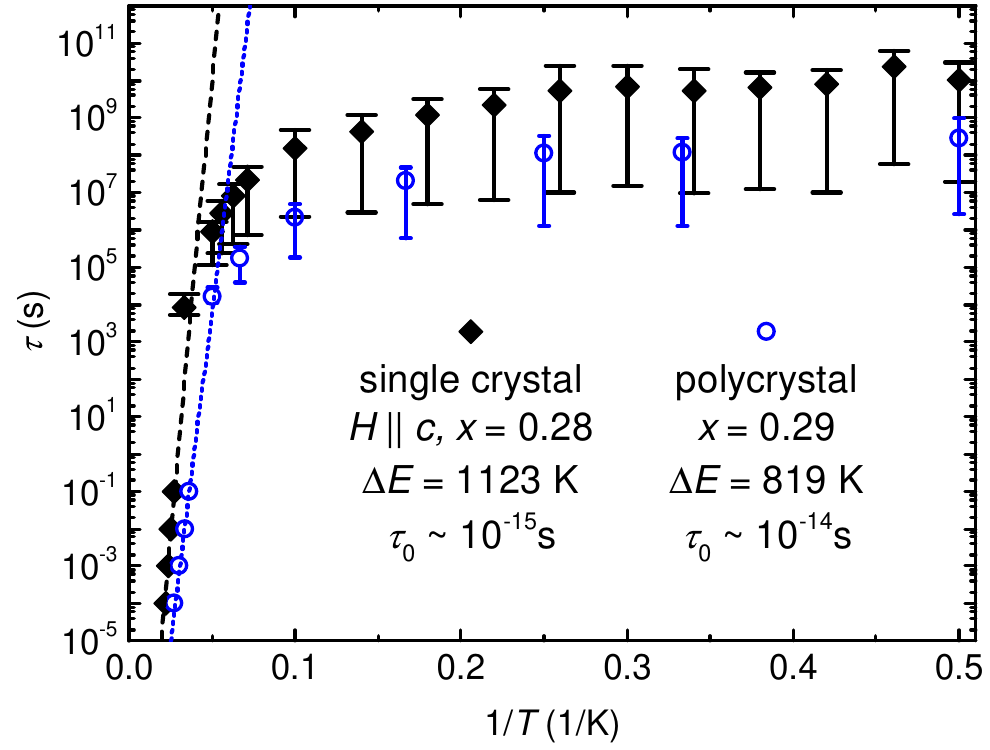}
 \caption{Temperature dependence of the relaxation time $\tau$ of Li$_2$(Li$_{1-x}$Fe$_x$)N as determined from ac susceptibility $\chi''(T)$ ($\tau < 1$\,s, bottom left) and stretched exponential fits to $M(t)$ ($\tau > 10^3$\,s). 
Thermally activated relaxation is indicated by the linear dependence of log($\tau$) on $1/T$ as observed for $T > 25$\,K  (Arrhenius behavior).
The corresponding effective energy barriers, $\Delta E$, were calculated from linear fits as shown by the dashed (single crystal) and dotted (polycrystal) lines. 
The plateau observed at low temperatures indicates the relevance of quantum tunneling for the relaxation process.}
 \label{arrhenius}
\end{figure}

The obtained values for the relaxation time $\tau$ are plotted in Fig.\,\ref{arrhenius} in the form of an Arrhenius plot.
For the single crystal, $\tau$ increases from $10^4$\,s to $10^{8}$\,s for decreasing temperatures from 30\,K to 10\,K. 
Below $T = 10$\,K the relaxation time increases only slightly and approaches a value of $\tau \approx 10^{10}$\,s.
Similar behavior is observed for the polycrystal with $\tau$ approaching a somewhat smaller value of $\sim 10^{8}$\,s.

\subsection{AC susceptibility}

\begin{figure*}
	\includegraphics[width=0.9\textwidth]{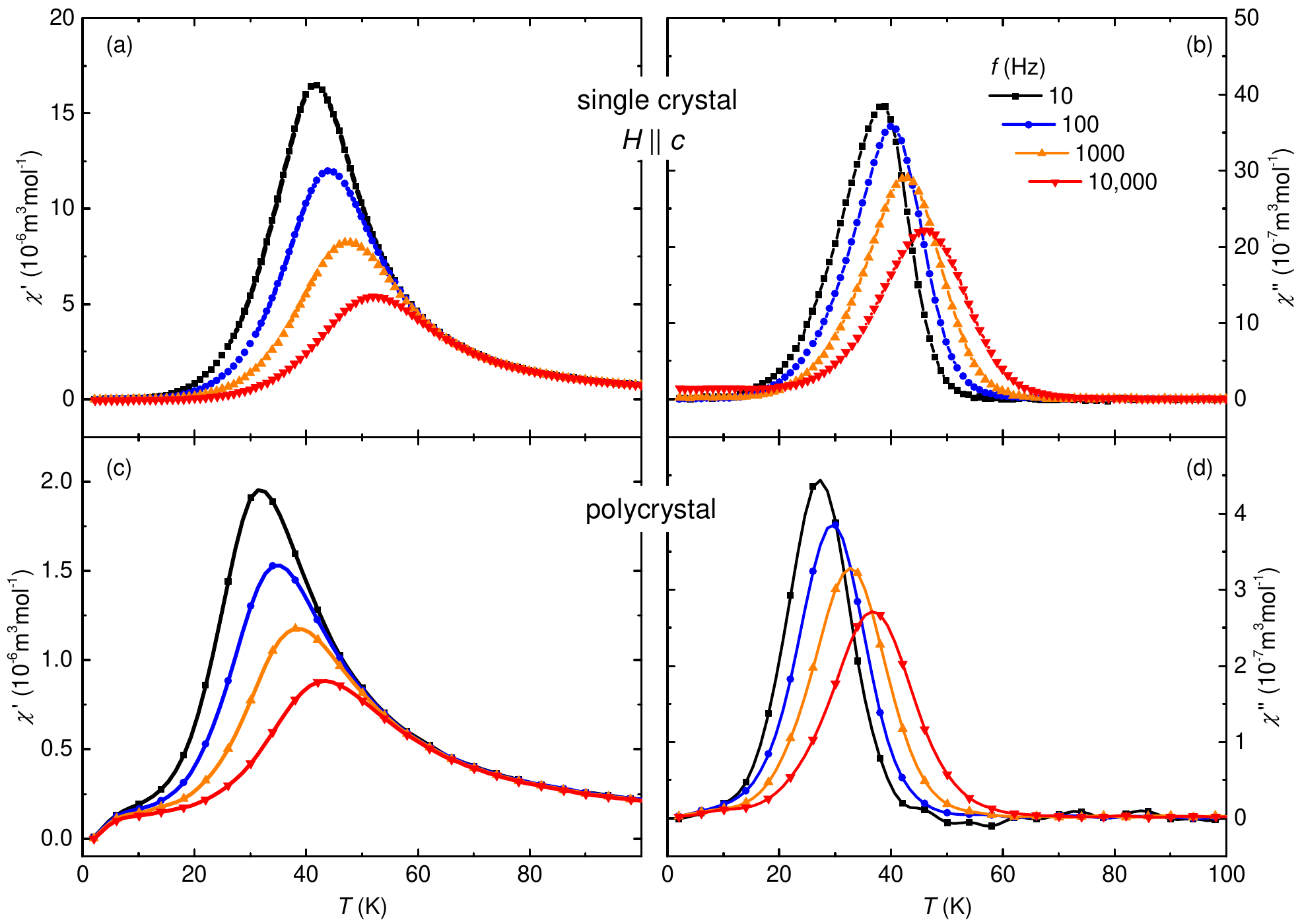}
	\caption{Temperature dependence of the in-phase, $\chi'$, and out-of-phase magnetic ac susceptibility, $\chi ''$, of Li$_2$(Li$_{1-x}$Fe$_x$)N.
(a,b) single crystal data for $H \parallel c$ ($x = 0.28$) and (c,d) polycrystalline pellet ($x = 0.29$). Lines are guides to the eye.}
	\label{ac}
\end{figure*}
For $T > 30$\,K the relaxation becomes too fast to be measured directly, however, $\tau$ is accessible via the imaginary part of the alternating current (ac) magnetic susceptibility. Fig.\,\ref{ac} shows the temperature dependence of the in-phase and out-of-phase part of the ac magnetic susceptibility $\chi'$ and $\chi''$, respectively.
The curves show a peak centered at $T_\mathrm{max}$, that shifts to higher temperatures with increasing excitation frequency (excitation field $\mu_0H = 1$\,mT).
A small but significant asymmetry of the peak in $\chi''$ with more weight on the low-temperature side is observed for the single crystal ($H \parallel c, \chi'' \parallel c$).

The maximum in $\chi''(T)$ corresponds to a maximum in the energy absorbed by the sample, meaning that the relaxation time and oscillation period of the ac field coincide: $\tau = 1/f$ at $T = T_\mathrm{max,\chi ''}$.
The obtained values for $\tau$ together with the relaxation times determined from direct measurements (stretched exponential fit, see previous section) are plotted in Fig.\,\ref{arrhenius}.
For $T > 25$\,K a linear (Arrhenius) behavior of log($\tau$) as a function of $1/T$ is observed with good agreement between direct and indirect measurements. 
A fit of $\tau(T)$ to $\tau(T) = \tau_0 \exp\{ \Delta E / k_\mathrm{B} T\}$ (dashed and dotted line) gives energy barriers of $\Delta E / k_\mathrm{B} = 1123$\,K and 819\,K for the single crystal and the pellet, respectively.
The estimated attempt times of $\tau_0 = 10^{-15}$\,s and $\tau_0 = 10^{-14}$\,s for single crystal and polycrystal, respectively, have to be considered as very rough estimates since they correspond to an extrapolation beyond the experimental data by more than 10 orders of magnitude.

The relaxation deviates significantly from Arrhenius behavior for cooling below $T \sim 25$\,K, where the increase of $\tau$ with decreasing temperature is much smaller than expected for thermally activated relaxation.
$\tau$ is essentially temperature independent below $T = 10$\,K indicating the relevance of quantum tunneling of the magnetization for the relaxation process.

Another characteristic quantity is the frequency shift of the maximum in the real part of the ac-susceptibility $\chi '(f,T)$\,\cite{Mydosh1993} . It can be quantified by calculating the value
\begin{equation}
 F = \frac{\Delta T_\mathrm{max, \chi '}}{T_\mathrm{max, \chi '} \Delta (\log f)},
 \label{eq:frequencyshift}
\end{equation} where $T_\mathrm{max, \chi '}$ is the temperature of the respective maxima of $\chi '(f,T)$ and $\Delta T_\mathrm{max, \chi '}$ refers to the difference between the highest and lowest $T_\mathrm{max, \chi '}$ of the investigated frequency range appearing in the denominator\,\cite{Mydosh1993}.
Calculating $F$ for Li$_2$(Li$_{1-x}$Fe$_x$)N leads to the values 0.07 and 0.08 for the single crystal and the pellet, respectively.

\section{$^{57}$Fe M\"ossbauer measurements}\label{moess}
\begin{figure}
\includegraphics[width=\columnwidth]{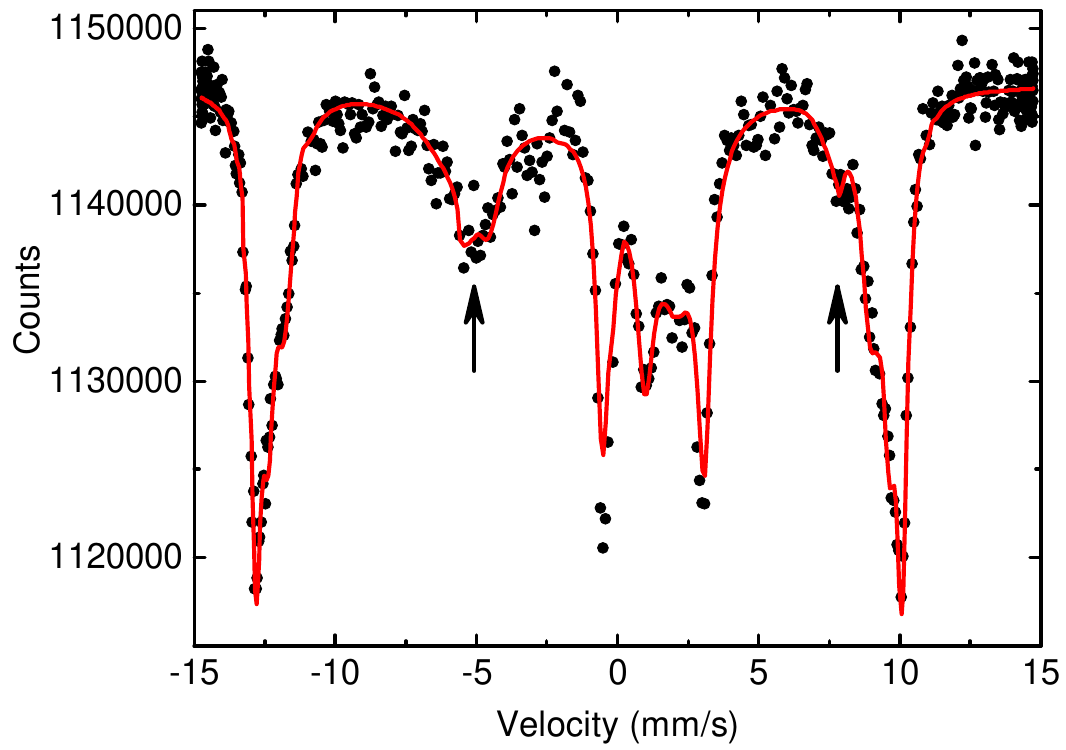}
\caption{M\"ossbauer measurement and transmission integral fit of Li$_2$(Li$_{1-x}$Fe$_x$)N with $x \approx 0.30$ at $T = 4.2$\,K. The arrows mark peaks with reduced intensity due to magnetic transition dipole selection rules for the $\gamma$-beam aligned parallel to the crystallographic c-axis.\label{moessbauer}}
\end{figure}

$^{57}$Fe M\"ossbauer measurements were performed on single crystalline Li$_2$(Li$_{1-x}$Fe$_x$)N with $x\approx0.30$ (nominal composition).
The sample was mounted with the crystallographic c-axis parallel to the $\gamma$-beam. 
Figure\,\ref{moessbauer} shows the M\"ossbauer spectrum obtained at $T = 4.2$\,K and the corresponding transmission integral fit using a static hyperfine Hamiltonian.
The complex spectrum with 6 main absorption lines in a wide velocity range between $-13$\,mm/s and +11\,mm/s proves the presence of static magnetic hyperfine fields on the timescale of the M\"ossbauer hyperfine splitting.

Consistent with previous experiments on polycrystalline samples\,\cite{Klatyk2002,Ksenofontov2003}, the total spectrum is described by seven subspectra that correspond to different lateral surroundings of the Fe site: 
species A with only Li as in-plane nearest neighbor ($n = 6$), species B with five Li and one further Fe ($n = 5$), species C with four Li and two Fe ($n = 4$), etc.
The measured spectrum is dominated by species A which accounts for 30\,\% of the area fraction. 
A combinatorial analysis leads to a probability of  $W_n = 6![n!(6-n)!]^{-1}(1-x)^nx^{6-n}$ to find $n$ Li atoms as (lateral) nearest neighbors\,\cite{Klatyk2002}. 
This yields a probability of $W_6 = 11.8\,\%$ for species A, significantly below the observed area fraction of 30\,\%.
A possible explanation is a tendency of Fe to avoid each other due to Coulomb repulsion giving rise to 'anti-clustering'.

For species A we obtained an isomer shift of $\delta=0.126(20)$\,mm/s, a hyperfine field of $B_{h}=70.21(20)$\,T and an electric field gradient of $V_{zz} = -156.11 (10)$\,V/\AA$^2$ in good agreement with previous results\,\cite{Klatyk2002}. 

Two of the peaks (marked by arrows in Fig.\,\ref{moessbauer}) should have zero intensity due to the magnetic dipole transition selection rules that do not allow these transitions for $\gamma$-beam $\parallel c$. 
The non-zero intensity is due to the finite mosaicity of the crystal, the presence of tilted crystallites and/or a misalignment of the surface normal with respect to the $\gamma$-beam.

\section{Specific heat}\label{5heat}
\begin{figure}
 \includegraphics[width=0.47\textwidth]{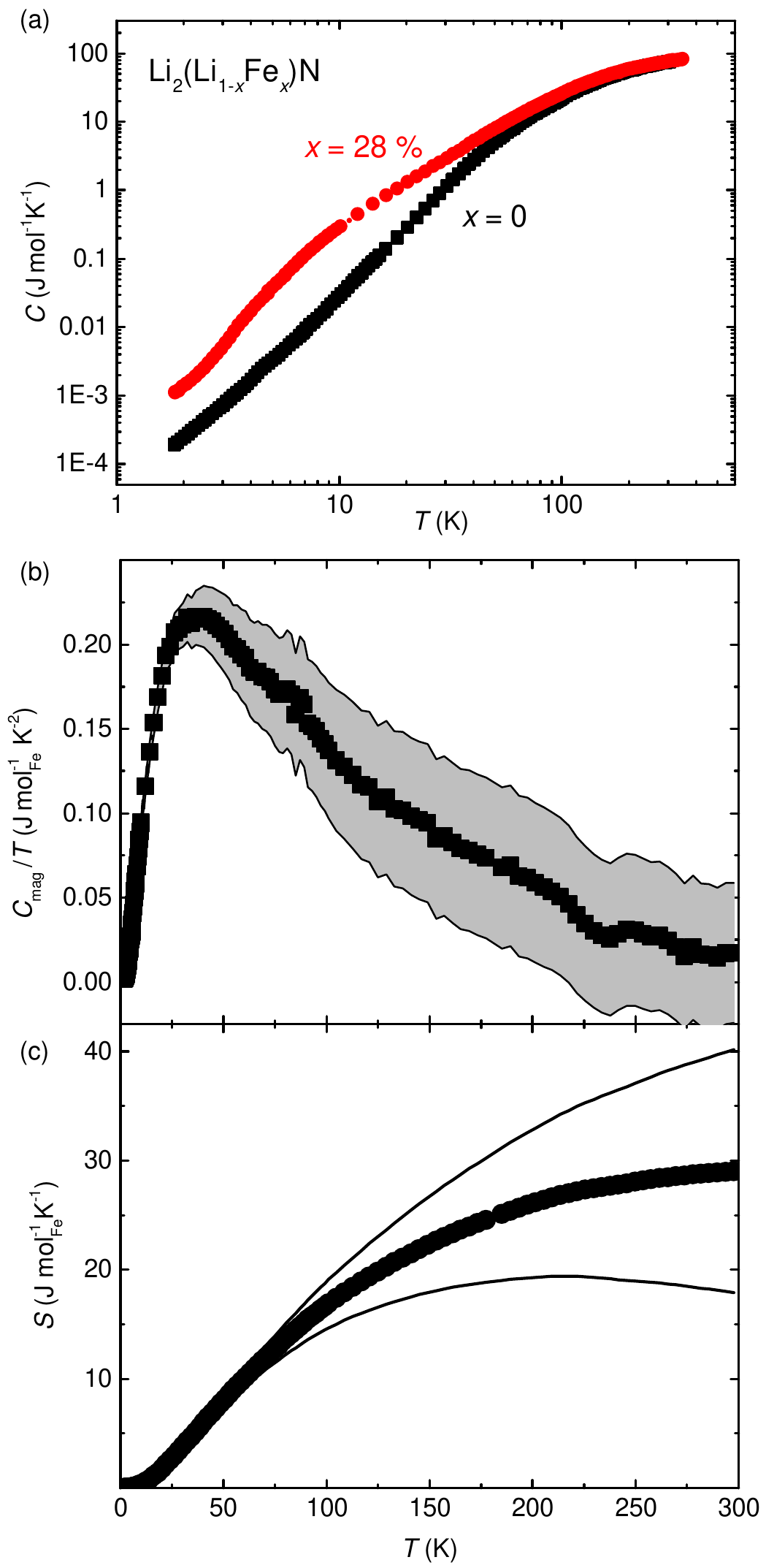}
 \caption{(a) Specific heat $C$ of Li$_2$(Li$_{1-x}$Fe$_x$)N single crystals with $x = 0$ (black squares) and $x = 0.28$ (red circles).
(b) Magnetic contribution $C_{\rm mag}$ calculated from the difference of the above curves (error shown by the shaded area, see text).
(c) Magnetic entropy obtained by integrating $C_{\rm mag}/T$ over temperature.
The solid lines indicate the upper and lower limit as given by the error of $C_{\rm mag}$.
} 
 \label{hc}
\end{figure}

Fig.\,\ref{hc}a shows a double logarithmic plot of the specific heat, $C$, as a function of temperature measured on two single crystals of Li$_2$(Li$_{1-x}$Fe$_x$)N with $x = 0$ and $x = 0.28$. 
In the undoped sample only phononic contributions are present with a decreasing slope towards higher temperatures approaching the Dulong-Petit limit of $C = 100$\,J\,mol$^{-1}$K$^{-1}$.
The room-temperature value of $C = 74.0(2.1)$\,J\,mol$^{-1}$K$^{-1}$ found for undoped Li$_3$N is in good agreement with previous results measured by adiabatic calorimetry\,\cite{Osborne1978} [$C = 75.22(15)$\,J\,mol$^{-1}$K$^{-1}$, $T = 298$\,K]. The Fe-doped sample shows the same value within the error bar [$C = 75.6(2.6)$\,J\,mol$^{-1}$K$^{-1}$]. 
For undoped Li$_3$N a Debye-temperature of $\Theta_{\rm D} = 670$\,K is obtained from the low-temperature limit.

For $T \lesssim 100$\,K, additional contributions become apparent in the Fe-doped sample that are attributed to the magnetic degrees of freedom of local magnetic moments of Fe. 
The difference between both measurements is plotted in Fig.\,\ref{hc}b as $C_\mathrm{mag}/T$ vs $T$.
The error was calculated by $\Delta C_{\rm mag} = \sqrt{\Delta C_{x = 0}^2 + \Delta C_{x = 0.28}^2}$ with $\Delta C/C  = \sqrt{ (\Delta m/m)^2 + (\Delta C_{\rm exp}/C_{\rm exp})^2 + (\Delta x/x)^2}$ using the following estimates: error of the sample mass $\Delta m = 0.2$\,mg, error of the heat capacity measurement $\Delta C_{\rm exp} / C_{\rm exp} = 2$\,\%\,\cite{Lashley2003}, error of the Fe-concentration $\Delta x = 0.02$. The resulting error of $\Delta C_{\rm mag}/T$ is indicated by the shaded area.
The magnetic contribution shows a maximum at $T \sim 35$\,K followed by a steep decrease for lower temperatures.
The shape resembles a broad Schottky anomaly rather than a sharp peak associated with a phase transition into a magnetically ordered state. 

Fig.\ref{hc}c shows the magnetic entropy $S_\mathrm{mag}$ obtained by integration of $C_\mathrm{mag}/T$. 
The curve approaches a value of roughly 30\,J\,mol$_\mathrm{Fe}^{-1}$K$^{-1}$ at room temperature and significantly exceeds the value of $S = \mathrm{R} \ln 2 \sim 5.8$\,J\,mol$^{-1}$K$^{-1}$ expected for entropy release from an ordered ground state doublet.
Estimates of the minimum and maximum entropy based on the $\Delta C$ shown in Fig.\,\ref{hc}b are given by the solid lines.

\section{Thermal expansion}\label{6dila}
\begin{figure}
 \includegraphics[width=0.47\textwidth]{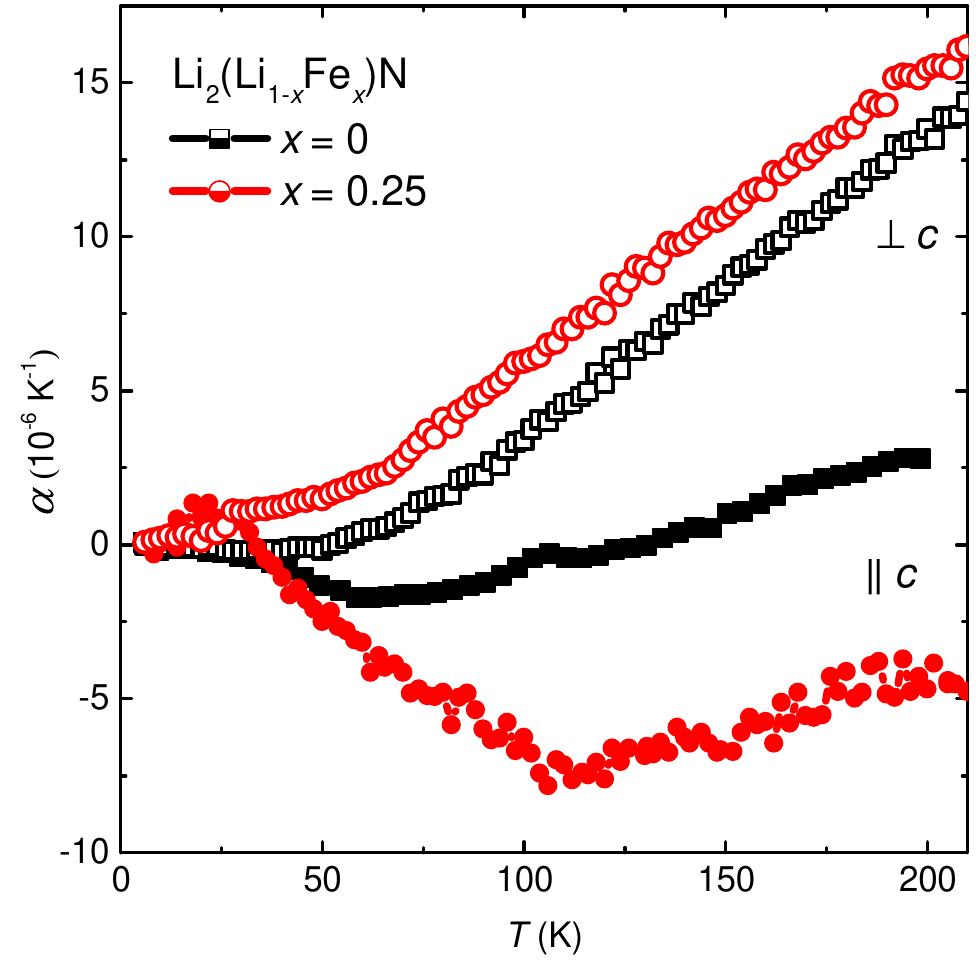}
 \caption{Thermal expansion coefficient $\alpha$ of Li$_2$(Li$_{1-x}$Fe$_x$)N single crystals with $x = 0$ (black squares) and $x = 0.25$ (red circles) as a function of temperature, measured parallel (closed symbols) and perpendicular (open symbols) to the crystallographic c-axis.}
 \label{alpha}
\end{figure}

Fig.\,\ref{alpha} shows the temperature-dependent, anisotropic thermal expansion coefficient $\alpha(T) = l^{-1}{\rm d}l/{\rm d}T$ of Li$_2$(Li$_{1-x}$Fe$_x$)N with $x = 0$ and $x = 0.25$ for $T$\,=\,5--220\,K. The measurement was performed upon warming with a sweep rate of 0.3\,K/min.
The thermal expansion perpendicular to the crystallographic c-axis, $\alpha_{\perp}$, is positive and increases with temperature over the whole range investigated for pure and doped Li$_3$N (open symbols in Fig.\,\ref{alpha}). 

In contrast, negative thermal expansion is observed parallel to the c-axis over a wide temperature range. 
For $x = 0.25$ a minimum with $\alpha_\parallel = -8\cdot10^{-6}$\,K$^{-1}$ is found at $T \approx 100$\,K. 
Upon further cooling, $\alpha_\parallel$ increases and changes sign at $T \approx 35$\,K. 
For undoped Li$_3$N, the observed values of $\alpha_\parallel$ are comparatively small. The local maximum at $T \approx 100$\,K coincides with the anomaly observed for $x = 0.25$ and has been reproduced on a second sample.
However, no anomalies were found in other physical properties in the vicinity of this temperature and its origin remains unclear. 

\begin{figure}
 \includegraphics[width=0.47\textwidth]{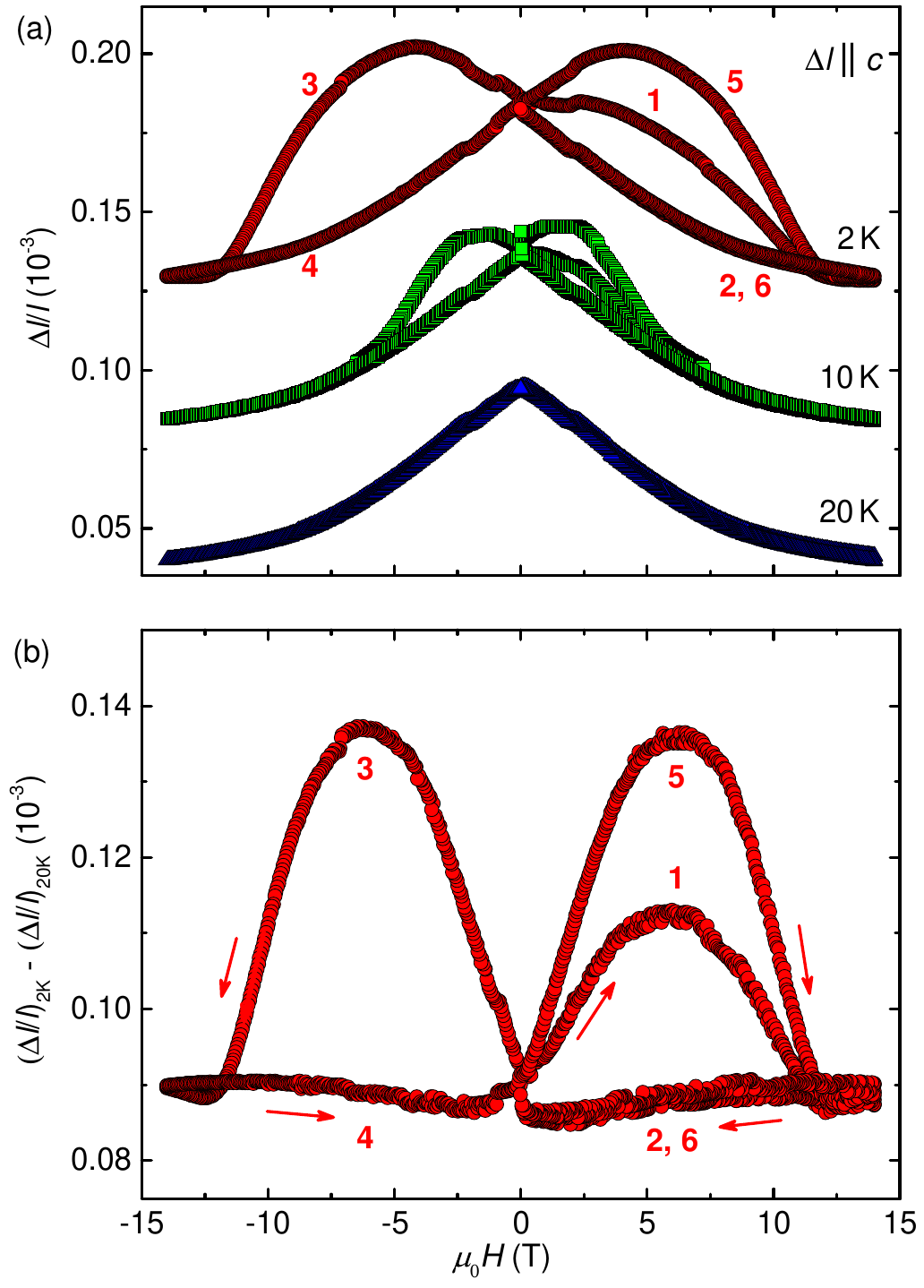}
 \caption{(a) Relative length change $\Delta l/l$ of a single crystal of Li$_2$(Li$_{1-x}$Fe$_x$)N measured along the c-axis as a function of a magnetic field applied parallel to $c$ ($x=0.25$, curves shifted for clarity). The numbers next to the 2\,K curve indicate the progression of the field sweeps. (b) Subtraction of the 20\,K curve from the 2\,K data shown in (a).}
 \label{deltaL}
\end{figure}

Figure\,\ref{deltaL} shows the magnetostriction of single crystalline Li$_2$(Li$_{0.75}$Fe$_{0.25}$)N. 
The isothermal length change as a function of field was measured along the c-axis for $H \parallel c$ (since hysteresis in $M$-$H$ emerges in this orientation). 
The sample was zero-field-cooled to the temperatures given in the plot ($T = 2$, 10\,K, and 20\,K).
Further measurements were performed at $T = 30$\,K, 40\,K, and 50\,K (not shown) and were found to be similar to the one at $T = 20$\,K. 
Small anomalies are observed for $\vert \mu_0 H \vert = 2$\,T in all loops at all temperatures measured and no relation to the magnetic states of Fe in Li$_2$(Li$_{1-x}$Fe$_x$)N is apparent. 
Between the isothermal loops, the temperature was increased to $T = 100$\,K and the magnetic field removed in order to demagnetize the sample.
The order of the field-sweeps is indicated by the numbers shown in Fig.\,\ref{deltaL} for $T = 2$\,K: beginning from the unpolarized state, $\Delta l$ decreases with increasing field up to the largest available field of $\mu_0H = 14$\,T (1). 
With decreasing fields (from +14\,T to zero), $\Delta l$ increases and reaches its initial zero-field value (2).  
For ramping the applied field to negative values, $\Delta l$ keeps increasing up to $\mu_0 H = -5$\,T, which is the field where the magnetization starts to decreases with increasing slope in the corresponding isothermal magnetization measurements (see Fig.\,\ref{m-h}). $\Delta l$ decreases with decreasing fields from $-5$\,T to $-14$\,T with a significant change of slope at $\mu_0 H \approx 12$\,T (3). 
When increasing the field from $-14$\,T, the hysteresis loop opens at $\mu_0 H \approx 11.5$\,T (4).
A symmetric loop is obtained upon further increase of the applied field to $\mu_0 H = +14$\,T (5).
Finally, the initial state is recovered after ramping the field back to zero (6).

For $T = 10$\,K, the width of the hysteresis loop decreases similarly to the behavior observed in isothermal magnetization measurements (Fig.\,\ref{m-h}).
No hysteresis is observed for $T \geq 20$\,K.
The overall behavior of the magnetostriction seems to reflect the isothermal magnetization $M$-$H$. 
All $\Delta l/l(H)$ loops cross at $H = 0$ since the thermal expansion does not depend on the direction of the magnetic moments (parallel or anti-parallel) to the c-axis for $H = 0$.
Similar behavior of the magnetostriction is observed in many ferromagnetic materials and can be accurately described by theory, see e.g. Ref.\,\onlinecite{Smith2003}. 
What seems more peculiar is the continued increase in $\Delta l$ after the sign change of $H$ (crossover from section 2 to 3 and from section 4 to 5 in Fig.\,\ref{deltaL}) that is absent at $T = 20$\,K.
$\Delta l(H)$ in sections 2 and 4 does not change significantly between $T$\,=\,2--20\,K.  
Subtracting $\Delta l/l(H, T = 20$\,K) from $\Delta l/l(H, T = 2$\,K) reveals a sudden increase at $H = 0$ whereas the resulting magnetostriction - which is directly related to the spontaneous magnetization - is constant in sections 2 and 4, as expected in the saturated state for $H \parallel M$ (Fig.\ref{deltaL}b).

In further contrast to the $M$-$H$ loops, the magnetostriction reaches saturation (in the sense that $\Delta l$ is reversible and comparable to the behavior at higher temperatures) from a zero-field-cooled state at $T = 2$\,K.
The magnetization, on the other hand, does not reach a constant value in the vicinity of $\mu_0H = -14$\,T (see Fig.\ref{m-h}).
The width of the hysteresis loop in $\Delta l(H)$ amounts to roughly 11.5\,T (on either side).
This value is in good agreement with the coercivity field of $\mu_0H_{\rm c} = 11.6$\,T.
Therefore, the effect of $H$ on $\Delta l$ diminishes as soon as the magnetic moment is no longer antiparallel to the applied field.
This indicates that the observed length change may be significantly affected by a torque acting on the springs of the capacitive dilatometer and does not fully reflect a change of the lattice parameter. 
An estimate of the magnetic energy in the 2nd quadrant of the $M$-$H$ loop (moment and field are antiparallel) and a comparison with the potential energy of the dilatometer spring supports this scenario: $E_{\rm mag} = - \mu B$cos($\sim$179.9$^\circ - \phi)$ and $E_{\rm pot} = k b \phi^2$  [$\mu$ := size of the moment, $B$ := magnetic induction, ($\sim$179.9$^\circ-\phi$) := angle between $\mu$ and $B$, $k$:= spring constant, $b$ := sample dimension perpendicular to the field].  
A small deviation from $180^\circ$ (misalignment) is necessary in order to overcome the unstable equilibrium for $\mu$ being perfectly antiparallel to $B$.
With $\mu = 4\,\mu_{\rm B}$, $B = 5$\,T, $k = 25.000$\,N/m\,\cite{Kuchler2012} and $b = 3$\,mm we obtain a minimum in the total energy for $\phi = 0.0015^\circ$ that corresponds to a displacement of $8\cdot10^{-8}$\,m.
This allows to explain the observed value of $\Delta l = 5\cdot10{^-8}$\,m (Fig.\,\ref{deltaL}, $l = 1$\,mm)

This effect may have been elusive in former magnetostriction experiments, since good-quality single crystals of hard magnetic materials do not show any appreciable magnetic hysteresis (e.g. Nd$_2$Fe$_{14}$B\,\cite{Sagawa1985} or $R_2$Fe$_{14}$B\,\cite{Hiroyoshi1986} with $R$ = Ce, Pr, Dy, Tm). 
Furthermore, the magnetic anisotropy energy and accordingly the proposed torque of $R_2$Fe$_{14}$B and related compounds is smaller than the one of Li$_2$(Li$_{1-x}$Fe$_x$)N\,\cite{Jesche2014b}.

\section{Discussion}\label{7disc}
Whether a material is a paramagnet with slow relaxation of the magnetization or a 'true' ferromagnet is not easy to answer.
Li$_2$(Li$_{1-x}$Fe$_x$)N shows characteristics of both.
On the one hand, the quasi-static coercivity (sweep rates lower than 10\,mT/s) at low temperatures even exceeds the values observed in LuFe$_2$O$_4$\,\cite{IIda1993, Wu2008}.
The only report on larger coercivity fields we are aware of ($\mu_0H_{\rm c} = 55$\,T) was measured on Sr$_3$NiIrO$_6$ in pulsed magnetic fields at sweep-rates larger than 25\,T/s\,\cite{Singleton2016}.
However, the values obtained in quasi-static fields are rather moderate ($\mu_0H_{\rm c} < 1$\,T)\,\cite{Mikhailova2012, Lefran2014}. 
We have also shown static magnetic ordering of Li$_2$(Li$_{1-x}$Fe$_x$)N on the time scale of M\"ossbauer spectroscopy. 
Furthermore, a ferromagnetic ground state is inferred from band structure calculations based on density functional theory\,\cite{Klatyk2002, Novak2002, Antropov2014}. 

On the other hand, a peculiar time-dependence of the magnetization is observed that is usually not associated with static ferromagnetism.
Even though ferromagnetic materials do show time-dependencies and relaxation effects, for example the {\it Time Decrease of Permeability} or the {\it Magnetic Aftereffects}, see e.g. Ref.\,\onlinecite{Ohandley2000, Cullity2009}, there are distinct differences to the behavior of Li$_2$(Li$_{1-x}$Fe$_x$)N as discussed below. 
In particular, we are not aware of any ferromagnetic material that shows a sizeable frequency-dependence of a maximum in $\chi ''$\,\cite{Mydosh1993} (at least not for $f < 10,000$\,Hz).
Instead, we have found magnetization dynamics reminiscent of superparamagnets, spin glasses and molecular magnets.
Just like each of these material classes are hard to distinguish from an experimental point of view\,\cite{Gatteschi2006, Mydosh1993}, Li$_2$(Li$_{1-x}$Fe$_x$)N cannot be simply attributed to either of these categories.

For spin glasses the frequency shift $F$ in the real part of the ac-susceptibility (see equation\,\ref{eq:frequencyshift}) typically assumes values of the order of 0.01 while superparamagnets and molecular magnets show $F > 0.1$\,\cite{Mydosh1993,Greedan1996,Gatteschi2006}.
Although the frequency shift of $F$\,=\,0.07--0.08 calculated for Li$_2$(Li$_{1-x}$Fe$_x$)N is only slightly larger than observed in insulating spin glasses, the characteristic sharp cusp in $\chi '$ at the freezing temperature (a sign for the collective freezing of the particles) is absent. Instead, we observe a rather broad peak in $\chi '$, typical for superparamagnets, where a size distribution of the magnetic particles induces a gradual blocking of the particles at different temperatures\,\cite{Bedanta2009,Mydosh1993}.
Furthermore in contrast to spin glasses\,\cite{Mydosh1993}, $\chi'$ does not approach a finite value for $T \rightarrow 0$ but instead decreases towards $\chi' = 0$ for decreasing temperature.
Also the ratio $\chi ''/\chi '$ is unusually large in comparison with spin glasses and instead more typical for molecular magnets\,\cite{Gatteschi2006}.

Another similarity to molecular magnets is the temperature independent relaxation rate approached for low temperatures (see Fig.\,\ref{arrhenius}) that indicates a quantum tunneling of the magnetization\,\cite{Novak1995,Friedman2010,Wernsdorfer1999b}. 
On the other hand, in comparison with molecular magnets we observe an intriguing combination of a huge effective energy barrier $\Delta E$ and a large relaxation time in the tunneling regime, together with a relatively high temperature at which the relaxation process deviates from Arrhenius behavior. 
Li$_2$(Li$_{1-x}$Fe$_x$)N shows a unique combination of energy barrier, critical temperature for quantum tunneling, and relaxation time in the tunneling regime (at the plateau).
We shall provide a comparison of these quantities with typical molecular magnets (a summary is given in Table\,\ref{tab:molecular_magnets}).

\begin{table}
	\linespread{1.5}\selectfont
	\begin{tabular}{l c c c c}
		\hline
		\hline		
		~ & $\Delta E/k_\mathrm{B}$\,(K) & $T_\mathrm{d}$\,(K) & $\tau_\mathrm{QTM}$\,(s) & Reference\\
		\cline{2 -5}
		Mn$_{12}$-acetate & $\sim 60$~ & ~$<2$~ & ~$\sim 10^7$--$10^8$ & ~[\onlinecite{Sessoli1993,Paulsen1995,Barbara1995}]~\\
		
		Fe(C(SiMe$_3$)$_3$)$_2$ & ~$\sim 325$~ & ~$\sim 20$~ & $\sim 1$ & ~[\onlinecite{Zadrozny2013}]~\\
		
		4$f$-based MMs & ~$>900$~ & ~25--50~ & $<1$ & ~[\onlinecite{Ganivet2013,Blagg2013,Gonidec2013}]~\\
		\hline
		Li$_2$(Li$_{1-x}$Fe$_x$)N & $>1000$ & ~$\sim 25$~ & ~~$\sim 10^{10}$ & ~this work~\\
		\hline
		\hline
	\end{tabular}
	\linespread{1}\selectfont
	\caption{Typical properties for three classes of molecular magnets (MMs) in comparison with Li$_2$(Li$_{1-x}$Fe$_x$)N. Given are the effective energy barrier for spin reversal in the thermally activated regime $\Delta E$, the temperature $T_\mathrm{d}$ below which a deviation from Arrhenius behavior is observed and the low temperature relaxation time $\tau_\mathrm{QTM}$, i.\,e. the relaxation time in the regime where quantum tunneling of the magnetization is observed.}
	\label{tab:molecular_magnets}
\end{table}

Mn$_{12}$-acetate, one of the most studied molecular magnets, has a relaxation time in the tunneling regime of $\tau_\mathrm{QTM} \sim 10^7$--10$^8$\,s\,\cite{Paulsen1995,Barbara1995} which comes close to the observed $\tau_\mathrm{QTM} \sim 10^{10}$\,s for single crystalline Li$_2$(Li$_{1-x}$Fe$_x$)N. 
The energy barrier is, however, much smaller in the case of Mn$_{12}$-acetate ($\Delta E \sim 60$\,K\,\cite{Sessoli1993,Paulsen1995} compared to $\Delta E > 1000$\,K). 
Also, the deviation from Arrhenius behavior is observed only for temperatures $T_\mathrm{d}$ lower than 2\,K\,\cite{Paulsen1995,Barbara1995}, i.\,e. much below $T_\mathrm{d} \sim 25$\,K observed in Li$_2$(Li$_{1-x}$Fe$_x$)N).

A particularly interesting material to compare with is the compound Fe(C(SiMe$_3$)$_3$)$_2$\,\cite{Zadrozny2013} since it shows the same structural motif as Li$_2$(Li$_{1-x}$Fe$_x$)N, namely the two-fold, linear coordination of the magnetic center. 
Although enclosed by an organic molecule (instead of an inorganic matrix), the magnetic center consists of only one single Fe-ion - as opposed to a cluster of magnetic ions - that is in the same oxidation state (+1) as the Fe-ions in Li$_2$(Li$_{1-x}$Fe$_x$)N\cite{Klatyk2002}.
The deviation from thermally activated relaxation is observed at $T_\mathrm{d} \sim 20$\,K\,\cite{Zadrozny2013}, which is relatively close to the value observed in Li$_2$(Li$_{1-x}$Fe$_x$)N.
Also the energy barrier, $\Delta E \sim 325$\,K\,\cite{Zadrozny2013}, is smaller but still comparable to Li$_2$(Li$_{1-x}$Fe$_x$)N. 
A major difference is observed in the magnitude of the relaxation time upon entering the quantum tunneling regime: $\tau_\mathrm{QTM}$ approaches values in the range of seconds\,\cite{Zadrozny2013} whereas $\tau_\mathrm{QTM} \sim 10^{10}$\,s is found for Li$_2$(Li$_{1-x}$Fe$_x$)N. Accordingly, Fe(C(SiMe$_3$)$_3$)$_2$ does not show any appreciable coercivity at quasi-static sweep rates.

As the last material to compare with we choose the Lanthanide ($4f$-) based complexes that show effective energy barriers as high as 938\,K\,\cite{Ganivet2013,Blagg2013}, which comes closest to the value observed in Li$_2$(Li$_{1-x}$Fe$_x$)N. 
Furthermore, the relaxation times also deviate from Arrhenius behavior at $T = 25-50$\,K\,\cite{Ganivet2013,Blagg2013,Gonidec2013} . 
However, similar to Fe(C(SiMe$_3$)$_3$)$_2$, the relaxation times in the low temperature regime do not exceed 1\,s\,\cite{Ganivet2013,Blagg2013,Gonidec2013}.

Another argument speaking against a collective (ferromagnetic) behavior in Li$_2$(Li$_{1-x}$Fe$_x$)N comes from a comparison of the 'ordering temperature' and the anisotropy energy. While in typical ferromagnets the anisotropy energy is several orders of magnitude smaller than the ordering temperature, the opposite situation is observed in Li$_2$(Li$_{1-x}$Fe$_x$)N (see section \ref{ch:magnetic_properties}), indicating that the hysteresis of the isothermal magnetization is not caused by exchange interactions, but by a slow relaxation of the magnetization due to the huge anisotropy energy.

\section{Summary}

Large hysteresis in $M$-$H$ emerges in polycrystalline and single crystalline Li$_2$(Li$_{1-x}$Fe$_x$)N at temperatures below $T \approx 50$\,K for field sweep-rates of some mT/s. 
Temperature-dependent specific heat reveals an entropy release significantly above the value expected for an ordered ground state doublet ($R$ln2). 
Thermal expansion and magnetostriction indicate rather low magneto-elastic coupling in accordance with slow relaxation of weakly coupled magnetic moments. 
Relaxation effects are observed in ac susceptibility and direct time-dependent magnetization measurements. 
For temperatures above $\approx 20$\,K those are dominated by thermal excitations whereas indications for quantum tunneling of the magnetization emerge at lower temperatures. 
Even though interactions between the magnetic Fe-moments are present for $x \approx 0.3$, the spontaneous magnetization seems to be a result of slow relaxation rather than collective ordering.

\section*{Acknowledgments}
We thank Alexander Herrnberger and Klaus Wiedenmann for technical support, Andrea Moos for performing ICP-OES and Robert S. Houk and Jenee L. Jacobs for ICP-MS measurements.
Michael L. Baker, Liviu Chioncel, Philipp Gegenwart, Theodor Gr\"unwald, Liviu Hozoi, Hans-Albrecht Krug von Nidda, Robert K\"uchler, Andreas \"Ostlin, Alexander A. Tsirlin, and Lei Xu are acknowledged for helpful comments and fruitful discussions. 
This work was supported by the Deutsche Forschungsgemeinschaft (DFG, German Research Foundation) - JE 748/1.

%

\end{document}